# Data fluidity in DARIAH – pushing the agenda forward

Laurent Romary, Inria& DARIAH

Mike Mertens, DARIAH

Anne Baillot, CMB & Inria

## Setting up a European Infrastructure – 10 years back

The history of DARIAH began in January 2006 when representatives from four European institutions[1] met to identify how they could join efforts in providing services to the research communities they served, with a strong focus on the humanities. The idea behind this meeting was to work towards a consortium of institutions, which would ensure long-term sustainability of the underlying infrastructure and a strong political voice vis-à-vis the EU. Each institution had a national role in coordinating or developing digital services in the humanities and could thus already speak with authority at that level.

DARIAH was thus put together as a top-down initiative of scientific information institutions, each having a duty to provide services to their respective research communities. In a way, one could say that DARIAH was conceived without the research communities themselves. But we would analyse this as a very beneficial factor since it gave us flexibility with regard to the actual scholarly coverage of our activities.

In the following years, the DARIAH model was completely inverted so that it has become a bottom-up organisation based upon active members and communities. In this context, there is a need at the highest level of the DARIAH management to reflect upon the complexity of the DARIAH landscape and identify the major priorities that are likely to be impacts on the scholarly communities in the future.

The present paper intends to be a contribution to this process by tackling a number of major issues related to the quality and status of digital information in scholarly work, and the possible role DARIAH could play in setting up and pursuing the corresponding agenda. After a quick presentation of DARIAH as it stands today together with some of the current achievements, we will focus on some concrete proposals that we would like to push forward in the coming period in order to achieve better services, availability, quality and scholarly recognition in relation to scholarly data sets. We would like to demonstrate that a series of parallel issues have to be dealt with in a coherent way if we want to be both successful in our endeavours and useful to researchers.

---

[1] Sheila Anderson, director of AHDS; Peter Doorn, director of DANS; Laurent Romary, director for scientific information at CNRS; Ralf Schimmer, representing Harald Suckfuell, in charge of scientific information for the Max Planck Society.

# A quick glance at the current state of DARIAH ERIC

## Missions and organisation

DARIAH has been set up at European level as an ERIC[2], the official European legal framework for research infrastructures, based upon a consortium of states that agree to support the infrastructure for a long-term period. In this context the main missions of DARIAH have been set up so that it would be an essential instrument to accompany the move towards digital methods in the arts and humanities. More precisely, DARIAH is responsible to achieve the following missions:

- Identify infrastructural needs of scholarly communities and provide support to go towards fulfilling these needs;
- Coordinate national contributions in establishing sustainable digital services for the arts and humanities;
- Contribute to the establishment of national infrastructural roadmaps;
- Participate to setting up the European agenda for infrastructures in Europe.

In order to achieve this, several organisational instruments have been set up in DARIAH:

- The general assembly validates the budget and the general organisation of DARIAH;
- The scientific advisory board accompanies DARIAH in setting up its scientific agenda;
- The national coordinator committee, with active representatives from all countries is the place where members compare and synchronise their national priorities;
- The joint research committee coordinates and synthesize the national contributions to offer concrete services for the community as a whole, in particular through its main instrument, the working groups, on which we will come back later.

This complex environment, which is needed to ensure maximum communication and concertation at European level, is managed by the DARIAH coordination office.

## A complex membership structure and a complex landscape

From the initial small group of countries represented in the first phase of DARIAH, the setting up of the ERIC consortium and its submission to the European Union have created an impetus towards more countries to join DARIAH, and the following fifteen founder member countries:

Austria, Belgium, Croatia, Cyprus, Denmark, France, Germany, Greece, Ireland, Italy, Luxembourg, Malta, The Netherlands, Poland, Serbia, Slovenia

In the course of the first year, two additional countries, namely Poland and Portugal, joined the membership, as well as five institutions form Switzerland as associated members. This last example reflects the fact that,depending on the national setting, actual participation in DARIAH can be a step-by-step process, as long as the various following essential components are in place:
- A strong network of academic institution that are willing to work together under a

---

[2]European Research Infrastructure Consortium.

clear leadership;
- Explicitly stated political support at Ministry level that ensures the (mid term) viability of the DARIAH participation from a national point of view.

As a whole, and without opening up here an in-depth analysis that could be the topic of a strategic paper in itself, we can observe that the DARIAH Membership is based upon quite a variety of models across the different countries, which in turn depend on the following factors:

- More or less strong national involvement in the development of digital services in the humanities depending on the actual existence of a national roadmap and thus capacity;
- Various sustainability models based upon a project based funding scheme or the establishment of national service centres that act as counterparts to DARIAH at European level;
- The balance between research and culture, depending on whether the preservation and dissemination of cultural heritage is seen as part of the DARIAH agenda or not;
- Last but not least, the actual development of research funding programs to foster digitally-based research projects.

This quick overview is in itself a good indicator of the challenges we face everyday in fulfilling our mission, still we shall see in the following section that we have already managed to achieve a significant number of results.

## First achievements

DARIAH is designed from the ground up to create and share resources for the digitally enabled arts and humanities, for the benefit of all. The chief of these resources are funded and sponsored on the national level, and submitted centrally as part of the formal membership contribution. Presently there exists therefore a vast library of tools, services, software, project outcomes, workshop proceedings, as well as work on standards and teaching resources. One of the principal aims of DARIAH is to promote and facilitate the re-use of such materials, in order to accelerate new research, enable a more responsive and open publishing environment for humanities data, allow collaborations to be concluded more efficiently, with the overall goal of supporting not just the digital services and complex digital object in arts and humanities research but to broaden and deepen the adoption of digital methods and digital workflows amongst humanities scholars. DARIAH is currently working on refinements to its collation of these so-called "in-kind contributions" from its members, in order to ensure very precise interrogation through a single portal. DARIAH is also in the position of revising the criteria for evaluating this wealth of resources, which will be touched on in terms of the proposed 'DARIAH Seal of Approval', below.

Another major advancement in DARIAH's operations recently, mentioned above as the 'bottom-up' approach, has been the establishment of working groups as the means, not only of offering defined fora for international collaboration but placing the research interests of the community that we serve at the heart of realising our strategic goals in DARIAH. This means that, in conjunction with our Scientific Board, DARIAH is extremely well plugged in to researcher communities, and hence the tools, services and resources that it is broadly defining and helping to deliver for the future are tightly allied with cutting-edge thinking in this area. The working groups will also be instrumental in making the 'DARIAH Seal of Approval' happen on a thematic level, with different aspects of the Seal (authentication standards, TEI protocols, etc.) being allocated to review and assent by the representative expertise of the appropriate working group. This will mean that those contributing to DARIAH will be able to concentrate their efforts in a particular area, be it authentication for example, or standards for annotation and publishing digital scholarly editions, knowing that the assessment will be undertaken by specialised academic peers within the overall progressive framework of DARIAH's goals and activities.

Additionally, DARIAH, both as an integrating and membership body, has also either centrally collated, facilitated or helped create a range of services,for application across a broad spectrum of humanities research. These extend from examples such as OpenATLAS[3], a database system for working with archaeological, historical and spatial data, to Ramses, an annotated corpus of Late Egyptian Texts, via Semantic Topological Notes (SemToNotes[4]), a topological image annotation and image retrieval system; GINCO[5], a software platform for the management and distribution of scientific and technical terminologies, and SYNTHESIS[6], an information system for the administration and promotion of cultural assets. In terms of this category of its "in-kinds" portfolio alone (there are 12 in all) the multifaceted DARIAH community has produced some 70 distinct platforms, services, tools, applications and software environments to diversify and deepen the use of new digital techniques amongst humanities scholars.

As can be seen from the above services, there is necessarily a strong and intrinsic bond between the kinds of research arts and humanities communities undertake, and primary collections. In this respect, certain of the larger-scale services are also fully-fledged DARIAH affiliated projects. These projects, which have co-arisen with DARIAH and its impact over the past few years but especially since 2012, have redeployed resources from the wider DARIAH setting and also the principles promoted by DARIAH centrally: Open Access, data-sharing, open standards, public humanities, and so forth. Having vision and a strong ethical commitment to open research in this area is as important to DARIAH as providing the practical and innovative means of doing digitally-enabled research. Thus EHRI as an affiliated project has worked in a novel way actively to bring the public into its development and the realisation of meaningful virtual relationships between highly dispersed and multilingual collections, and Cendari[7] has not only repurposed DARIAH services, such as NERD for manuscript annotation but has consistently worked towards researchers and collection holders making their work more open.

One of the domains where DARIAH has been particularly active since even before its official start is that of training scholars in digital methods. Under the impulse of the Virtual Competence Centre dedicated to research and education (VCC2), but also in conjunction with a number of DARIAH-affiliated projects, various activities have led to concrete support for humanities scholars, especially where they are having initially to be self-taught in digital methods:
- Numerous summer schools have been organized in the past five years targeting a broad range of scholarly communities (historians, archaeologists, philologists) or covering general aspects of the digital humanities methods from an editorial (publication) or technical (data representation, programming) point of view;

---

[3] http://www.openatlas.eu
[4] http://hkikoeln.github.io/SemToNotes/
[5] https://github.com/culturecommunication/ginco#what-is-ginco-
[6] https://www.ics.forth.gr/isl/index_main.php?l=e&c=271
[7] http://www.cendari.eu

- A reference Course Registry[8] has been set up and put into production, which gathers descriptions of the multiple curricula in European higher education institutions that are open to humanities students and cover the skills needed in and for the digital humanities domain;
- Two flagship projects have received support at European level in this area. The first of these is DiXiT[9]. The Marie Curie Network has here coordinated the most prominent centres in the domain of digital scholarly editions and has allowed numerous young scholars to benefit from a large spectrum of expertise from across Europe. More recently, the Erasmus+ program supported the #dariahTeach[10] project, which compiles and creates re-usable (and open access!) teaching materials for the digital arts and humanities.

The strong interest in and support received for these activities makes it clear that one of the many futures for DARIAH will remain intrinsic to transmitting expertise within the digital arts and humanities.

With this general presentation of on-going activities and results, we hope we have conveyed the general dynamism that exists within DARIAH. Still, there is a sense for most of us that DARIAH should have a vision of where it should impact the scholarly communities in priority. The rest of this paper is intended to describe some component for such a vision, as we intend to implement it in the coming phase of DARIAH.

# Ensuring data fluidity in the arts and humanities scholarly community

Even if it is an obvious point, DARIAH is, as a digital infrastructure, a data-centered one, and can thus be differentiated from more physical infrastructures such as GEANT, EGI or the forthcoming E-RIHS (for cultural heritage analysis equipment). As a consequence all DARIAH activities, from training to the deployment of technical services, are targeted at facilitating the availability and exchange of relevant data for research in the arts and humanities. We have already described a number of consequences of this data-centred strategy for DARIAH in [Romary, 2011] and so the intention here is to suggest and outline areas in which DARIAH should be particularly active in the short and mid-term, in order to deliver concrete benefit for and at the service of scholarly communities.

## Hosting as a priority

An essential pillar of the open access movement in the last decades has been the setting up of a network of publication repositories, which has allowed various bodies (cf. [Romary & Armbruster, 2010] and [Armbruster & Romary, 2010]) to offer strong services for the dissemination of scholarly papers online, with rigorous technical and editorial conditions. In some communities, such as Computer Science or Physics, this has facilitated a shift from the individual dissemination of scholarly content though web pages (and before that, private snail mails to colleagues) to a more coordinated approach that, among other things, has secured the preservation and long-term access to the corresponding content. There could be many lessons to be learnt from this long history of the publication repository landscape, but we can at least identify how much the specificity of the scholarly object "paper" has had an impact at various levels on the way such repositories have been conceived.

---

[8] https://dh-registry.de.dariah.eu (see also http://dariahre.hypotheses.org/218)

[9] http://dixit.uni-koeln.de

[10] http://dariah.eu/teach/index.php/2015/05/21/welcome-to-dariahteach/

If we now come back to the issue of making the dissemination of scholarly data in the arts and humanities more fluid, we cannot but observe that the main request that scholars express when they produce research data in conjunction with their research projects is to be able to identify a clear setting where such data can be hosted in a trusted way.

In a way, scholars' expectations are at odds with what the landscape looks like at present, seen from the point of view of infrastructures. There is indeed a wealth of possible repository solutions. First there are a few European countries that have set up generic solutions for hosting research data in the humanities. Repositories such as Nakala[11] in France or EASY[12] in the Netherlands already offer robust services and have integrated thousands of datasets since their respective launches. There are even completely generic solutions such as D4Science[13], used for instance within the Parthenos project, that offer to record any kind of scholarly output from publications or reports to datasets in any scientific domain. On the opposite side of the spectrum, we see that highly specialized hosting possibilities exist in relation to initiatives that intend to serve specific object types or scholarly communities. In this category, we can name services such as MediHal[14] for images (for all scientific fields); infrastructures dedicated to the management and/or hosting of digital editions, such as TexGrid[15] (covering the whole editorial workflow) or Tapas[16] (for TEI documents), or generic infrastructures dedicated to linguistic content such as Ortolang[17], but which could as well host complete documentary corpora.

One of the essential priorities for DARIAH will be to be able to find the optimal compromise between generic and specific deployments for data repositories, but also to be able to deal with the complexity of polling together hosting services at European level. There is indeed a major challenge in making sure that hosting platforms do not reflect the fragmented national picture in Europe whereby countries will be reluctant to host (and thus to pay for) data that is not issued by scholars from its own research community. We thus need to find way, and probably business models, so that data hosting becomes transparent for DARIAH users while making sure that there is a fair and balanced distribution of costs. As addressed already in [Blanke 2016] linking crowds (scholars) and clouds (host) will be the key to the generalized dissemination of digital content in the humanities.

---

[11] https://www.nakala.fr

[12] https://easy.dans.knaw.nl

[13] https://www.d4science.org

[14] https://medihal.archives-ouvertes.fr, based on the same principles and infrastructure as HAL national publication repository platform

[15] https://textgrid.de

[16] http://tapasproject.org

[17] https://www.ortolang.fr/

## Towards a DARIAH seal of approval

The notion of a DARIAH Seal of Approval[18] could be seen as tending towards being essentially repositories-centric, given that the majority of specifications in this general modus often relate only to data the management of data per se, as requirements to be met, so as to comply with any technical demands of repository services. We think we need to go beyond this necessarily limited operational view and offer instead a set of broader reference features that may apply directly to the data sets and other research outputs themselves, and that will simultaneously integrate required production and provenance values (related to whom has produced it, where and under which conditions it is hosted) but also fundamentally spell out how the data as such could be further and reliably re-used and redeployed in entirely new research contexts. Thus we aim to go from a repository-centric view of data to a scholarly-centric one.

The challenge for humanities research in general in this area straightforwardly expressed: how to find the sweet spot between taking advantage of generic and therefore highly reusable digital tools, software and components on the one hand, and maintaining a necessarily high degree of academic freedom to pursue novel and hitherto inconceivable research questions these digital affordances altogether allow, on the other. In short, how can one systematically foster innovation? The meme of the laboratory has been in vogue within especially library and archival circles in recent years to signal data-driven and open humanities research[19], and it remains a powerful and apposite analogy. A lab without ideas to be tested is sterile, yet to be operational, a lab has to be rigorously maintained and work against known quantities and standards. It is in this spirit that the DARIAH Seal of Approval should be read: designing coherence for and at the service of experimentation.

The DARIAH Seal of approval would work in two main directions – to acknowledge that tools, services and software produced by the DARIAH community would meet criteria that allowed for their greatest potential reuse, on the one hand, and certify those primary collections *at item level* with which researchers wish to work as capable of maximum enrichment and subsequent access. There would be a number of discrete areas, modelled on the current in-kind contribution categories that DARIAH allocates to its member submissions to the central repository of shareable resources.

Digital objects are used in a wide variety of different contexts. Therefore quality here relates, not to absolute values but to use values. The Seal of Approval would not so much require of collection holders that certain minimum standards as such would need to be maintained as that information about the appositeness of any digital object be clear, precise and available to be exported or harvested. Since minimum standards at a technical level change over time, with the introduction of more powerful hardware and software, these would in any case have to be revised. But holding to some basic principles (describing a collection or items within any collection as being available via a well-documented API, what programming languages had been used for any tool or service under which digital objects were being made available) would greatly enable scholars to scope, find and identify resources that would work within the framework of their own particular projects and research questions with far greater surety. Such clear statements would also make it far easier for academics being asked to do so, to 'give back' their enrichments or annotations in a manner that could more readily be taken up and utilised by the CHI in question.

---

[18] We are obviously elaborating here on the successfulData Seal of Approval (http://datasealofapproval.org/en/) developed, among others, by our colleagues from DARIAH at DANS in the Netehrlands.

[19] See https://dariah.eu/activities/dariah-theme.html

These would not be classic data curation standards (which it goes without saying would also be applied) but beyond data conformance we need to focus now on data *performance*. The fact is that re-usability is sustainability. Services and collections will only last as long, not as there a need for them but to the extent that their use in new modes of arts and humanities research is smooth and unproblematic – and for this the knowledge that academics need is not so much technical as logistical and legal: can it be downloaded? What are the licence terms? Is it available in raw text, json, XML, has it been transposed to RDF, etc.? These conditions need to be applied not just to wholesale collection metadata descriptions beyond the data in question but intrinsically as part of the data itself. These logistical considerations as well as the technical ones, for data performance and conformance, will also be an element of a further DARIAH-related initiative, PARTHENOS, which is designed to ensure the integration of outputs from both CLARIN-ERIC and DARIAH-ERIC, and that a workable, overarching framework is correspondingly put in place. This all being said, what are the kinds of actions that the DARIAH Seal of Approval will presuppose?

### The Autonomous Data set

As stated in the above, data conformance has been tied very largely to the needs of data containers, not data users, i.e., data contexts. In that sense, we need to move towards describing data in ways that make them independent of the service architecture or maintenance routines in which they find themselves. But what qualities would an autonomous data set have? How would it operate? There are some illustrative examples or prior suggestions of the qualities that autonomous data should have. In terms of curation, Manfred Thaller has already extensively described what he terms self-preserving objects, data that has "information (added) to the object, as is required to make it fit for processing on radically changed platforms within radically changed environments" (Thaller, 2013). Another example of data descriptors, and hence potentially greater meaningful access and reuse of digital resources, that are based not on procedural or technical parameters but on the intrinsic qualitative context of the data and objects involved, is CAMELOT, developed at the Oxford Digital Library[20]. The categories of meaning in this data model would allow relationships between objects and their future re-use environments to be ascertained. So, not only are identifiers such as certain place types (academy, library) and person available within the model but aspects such as "Research funding, administration and projects, academic institution structure, scholarly activities, research communities, creative works, manifestations, instances, collections and aggregations...[and]...annotations". These are the very types of scholarly-centric notions that should be applied to data as much as repository-centric ones, and we would want to encourage and embed as part of the scheme.

### Authorship, provenance, reuse and citability

At the level of the monograph or article, authorship is a relatively unambiguous matter; despite some challenges, even the use and impact of articles with multiple authors can be overcome, despite such considerations as researchers moving between institutions due to career decisions, and research assessment value can still be properly allocated personally and institutionally. However, in the scenarios we are highlighting, with CHIs providing primary resources that researchers may build on or add to qualitatively, the research outputs may consist of annotations, other object enrichments, or algorithmic and code improvements. Additionally, if we imagine a much greater reciprocal arrangement between researcher and CHI, whereby the CHI gains greater recognition for the material that has been (often painstakingly) made available, the argument could be presented that in this context provenance equals authorship. Also, there remain issues around consistently and systematically being able to trace the trajectory of general archival use and access to the precise research impact of particular resources, especially where archives or special collections are offering new formats, such as video[21]. The opposite situation would have a positive sustainability effect on CHIs; the more it would be possible to demonstrate a clear relationship between use of digital (and analogue) archival resources, their citation and their formal research assessment and altmetric impacts, the easier it would be for CHIs to make the argument for funding based on evidence. This is crucial in a climate and at a juncture in digitally-enabled arts and humanities research, where to an extent CHIs would ideally want to take in researcher enrichments of primary digital resources,and researchers conversely are looking for ways of ensuring the long-term maintenance of their digital enhancements.

---

[20] http://camelot-dev.bodleian.ox.ac.uk

Overall then we would require two basic things: on the one hand, much greater formal acknowledgement of and a way of auditing and assessing research outputs beyond the format of the book and article, so as to encourage amongst especially early career researchers further and more varied interactions with digital primary resources. On the other hand, a means of recognizing the 'authorial' hand of CHIs in facilitating access to and, in eventually drawing in researchers' digital enrichments, the conceptual and material enlargement of collections for further (re-)use.

There is also one more potential category of citability, and this refers to one particular, by now, highly dynamic aspect of digital object enrichment – user interactions, which, in their available, amalgamated form, are also a rich source of material in their own right for arts and humanities scholars. How would one cite a particular interrogation of such user-interaction data in a persistent way, such that it could be invoked reliably, thus bringing arts and humanities research closer to the overall research quality not simply of quantification but of reproducibility? As well as conformance to extant international formats for data referencing that should also become a significant aspect of the DARIAH Seal of Approval, there is also the potential for DARIAH to develop and set some new ones for citability.

**Access**

According to the above, therefore, conditions of access to the documents and collections would necessarily take into account the huge variety of data and metadata available through CHIs involved in these research processes and actively contributing therefore to research outputs. The CHI part of the Seal of Approval would mark out therefore significant ground for participation in the scheme. Regarding metadata, free (as in speech and beer) access to document metadata would be made available (including document enhancements). In terms of images, access to scans of any document or image according to current quality standards would be granted on upon request. User interaction data would also be made available on request. The chief aspect governing this part of the Seal of Approval would therefore be licensing

---

[21] For example see [Sinn, 2012] and http://www.sr.ithaka.org/wp-content/uploads/2015/08/supporting-the-changing-research-practices-of-historians.pdf

### Licensing

In order to be eligible for any level of the DARIAH Seal of Approval, licence statements would have to be specified at all times and explicitly for each unit of the available content in question. The terms under which objects and metadata would be licensed to academics belonging to institutions that themselves were organisational signatories to the DARIAH Seal of Approval would need to be defined precisely from the outset and so designed so as not to be susceptible tochange. For this reason, it is recommended to implement from the beginning licences that will be as open as possible, which would prevent downstream use and re-use more restrictive at a later point.

Default licences for CHI content, with further options depending on level of openness versus level of 'recognition reward' within the DARIAH Seal of Approval' scheme desired (Bronze, Silver, Gold) would be described and recommended ; similarly there would be default licences for enrichments. A neutral scale might be, rather than stipulate only certain instances of known concrete licences, 'not-stated', 'proprietary', 'restricted', 'open'. Concrete and current licence options could then be mapped onto these categories. However, the standard licence to be used would be an open one, with exceptions having to be applied for (and which naturally can be entirely legitimate, in the case of personal data and constraints placed on the use of material by depositors).

### Technical setting

Finally the guidelines shall contain some basic technical principles, which may apply either to the repository at large, a particular collection, or at the item level. As in the case of the Data Seal of Approval, features or exact technical requirements or expectations will not be made[22]. Instead, statements such as:"Is the data represented and stored in formats that are compliant with international standards" or "Is the data optimally represented for long-term legibility of the content", or "Does your digital content have PIDs" would be the kinds of guiding questions the DARIAH Seal of Approval would prompt in aspirants, to lead them through a series of realisable expectations that would allow institutions to apply for a level of assurance in the re-use of their digital collections that would be both achievable but also according to broadly acceptable and established practices. Again, it is possible to imagine that beyond the Seal of Approval guidelines, a decision tree tool could be made available that would enable institutions both to correctly establish the level of re-use impact for their collections they would want to achieve, the requisite types of technical, logistical or licensing features their collections should intrinsically possess, and therefore what they would need to do in order to attain one or other of the levels of scholarly data assuredness within the DARIAH Seal of Approval.

Conversely, research institutions would be asked about what level of academic commitment they wanted to achieve; however, as in the case of freely available metadata as a basic requirement of CHIs, there would be fundamentals to be met by research organisations wishing to comply with, and therefore obtain recognition from, the scheme. One of these would be an unerring commitment to offer any enrichment also back to the CHI that had provided the original digital resource on which they are based. The CHI institution would not be obliged to accept but an offer should be made, which could still be taken up in a sense, even where the local technical capacity might not exist immediately, through the provision of pointers to a reference host elsewhere. Another stipulation for acquiring even the basic level of scholarly data assuredness on the part of researchers/research organisations and CHIs would be the consistently applied use of extant authority files, whether these be for places, persons, companies, and so forth.

---

[22]See http://datasealofapproval.org/en/information/guidelines/

## Data re-use charter

As important as it is, stating clear requirements regardingdigital data and being able to offer solid hosting services to researchers still remains only one component in a wider landscape of data exchange mechanisms. The major issues at stake have to be addressed not as isolated aspects, but as a whole. With this objective in mind, we introduce in the following section a course of action aiming at moving ahead swiftly at the service of scholars, but also of cultural heritage and other institutions that act as service providersto the whole community.

Taking a look at the fundament of arts and humanities research, we see that it is mainly based on the analysis of what could be called more generally'human traces'. This concept covers a variety of possible concretions: artefacts, works of art, written documents of all sorts, recordings etc. All of them have a historical dimension in common: they are inscribed in a (transmission, preservation) tradition. They also have in common to begenerally hosted in cultural heritage institutions, which themselves can be of various status and importance, ranging from national libraries and archives down to small, local museums endorsed by regional or city administrations. These institutionsare not all bound to an identical organisational or institutional setting; on the contrary, they can fall within various domains (public, private, foundations etc.).

Scholars exploring collections or exploiting single documentsfromCHIs all face similar and recurring problems when it comes to what they can do with the material they identify as relevant for their research. There is no generally valid rule as to how much they can quote, duplicate and furthermore republish in their scholarly work. This question extends to thevarious forms of researchdissemination bound toarise from this work and which range from traditional publications to complex productions yielded from the institution's content: catalogues, archival research guides, collections of images, transcriptions. From a cultural heritage point of view, it is not even clear what status such productions would have, even if, for instance, some institutions were enabled and encouraged to host the researchers' by-products insofar as they may be seen as complementary resources to the original material.

The lack of a clear and comprehensive framework that could serve as a general baseline for interactions between scholars and cultural heritage institutions is a hindrance to both the development of further research projects based on highly valuable documentary collections, and the visibility of the institutions themselvesas key actors in the research ecosystem.

The present initiative, launched by DARIAH-EU, but which aims at being widely inclusive to all organisations related to cultural heritage institutions,will provide such a framework. It is conceived as a win-win setting that has the strategic potential to act as a reference for all interactions between scholars and CHIs. It will be capable of relieving individual scholars or collection curators from the challenge of having to rule again and again on how to use and re-use documentary material on a case-by-case basis.

### Relevant use cases: challenges and options

In the course of finalising the charter's scope and phrasing, we need to gather relevant use cases that are bound to help us make sure to cover the majority of the scholarly community's needs. We will only present here two standard cases that reflect the kind of hurdlescommonly experienced. They should be enriched with further analyses, solutions and comparison with further use cases in a close future.

### Re-using iconographical material in publications

A typical example has occurred in the context of finalizing the publication of the archival research guides[23] produced in the context of the Cendari project at the beginning of 2016. These guides have been mainly written by scholars working on the medieval and First World War periods in the context of the transnational access program which allowed young scholars to dive into one specific theme for the corresponding period and gather historical evidence on the basis of the archival data made available to them through the Cendari architecture.

Although instruction had been given to them to check and quote the appropriate source for any iconographical material they would use in their research guides, we came across the situation where the actual status of the illustrative images was just unclear. In emergency, some further checks had to be carried out and at times images were taken out from the publication. In a way, the Cendari project was left in a situation where no clear intellectual copyright clearance could be activated for-reusing even a single image taken, for instance, from the web site of a cited archive.

In this case, we have to deal with a paradox: the commercial publication domain for instance has already set up a scheme that accounts for the "re-use of limited amounts of material from published works"[24]. The STM permission guidelines [STM 2014] state for instance the number of figures and tables that can be re-used in a publication without requiring any specific permission. It is noticeable too that some publishers also require to be notified of such use in any case.

So why should scholars, and infrastructures, spend so much time in checking up such simple rights, at the price that material is not actually re-used if a doubt remains concerning the corresponding rights?

### Digital edition from a physical textual source in a CHI

For many scholars studying primary textual sources, the production of an edition is an incompressible part of the research process. Digital editions reflect and accompany the evolution of this dimension of the research process.

Over the past years, the standards for digital scholarly editions have evolved towards a greater inclusion of the documentary basis within the editorial work itself. Many digital editions now comprise the transcription and (fine) annotation of a corpus of documents gathered from one or several libraries or archives – which are likely to become part of the edition itself. At each stage of his/her activity uncovering, unlocking, editing and exploiting these sources, the scholar is likely to face the following challenges and questions:
- Looking at potential sources, and beyond the kind of searches one could make on a portal such as Europeana, s/he would need to know whether, for the CHI s/he is interested in, there exists a digital catalogue of the collections and items, and whether s/he is allowed to take partake of the totality of any related information for future use or publication;
- Once s/he has identified required items in any library or archive, how can s/he be aware of the existence of digital surrogates or whethers/he is allowed to make and publishphotographs or scans;
- To what extent is the scan quality made available by the CHIs compatible with scholarly standards? When scan quality has improved noticeably, how can a researcher update his/her edition with regard to the scan quality (which requires the CHI to realize new scans and let the researcher use them according to the same terms as it was the case before), how can s/he negotiate the long-term cooperation with the CHI in this perspective?
- How to merge archival, librarian and research contribution to the metadata in a common dataset that makes the contribution of each partner visible and is quotable (question of multi-institutional enrichment, of hosting, of authority, of format compatibility and of the value of rich metadata in term of scholarly recognition);
- Last but not least, how much can the source material be actually re-used within an online publication or even a printed object, as is still the case for some scholarly editions.

---

[23]see http://www.cendari.eu/thematic-research-guides/intro-thematic-research-guides

[24]http://www.stm-assoc.org/copyright-legal-affairs/permissions/permissions-guidelines/

As we can see, and our list is probably not exhaustive, the level of potential complexity we reach here goes way beyond what an isolated scholar could decently understand and deal with. There is thus an emergency to provide a clearer setting for such kinds of scenarios.

## Which actors for which partnership?

To pave the way towards the design of a common and generic data re-use charter for cultural heritage content, we first need to see what this would mean for each major category of stakeholder and in what form they could engage in such an endeavour, taking into account costs and benefits. The orientations sketched below are designed for the three main categories of stakeholders, namely cultural heritage institutions, scholars and hosting services.This analysis and the action lines drafted here are conceived as the basis for a wider dialogue that could and should enrich, andadd greater precision to, what we suggest here.

### Cultural heritage institutions

Cultural Heritage Institutions can be very diverse. History, mission or focus vary from one institution to the other, often involving to consider strong regional specificities and legacy cultures, not to mention the impact of history. The charter aims at considering them in the generality of their function as curators of collections and objects in their physical form and as potential primary initiators of corresponding digital surrogates, from basic descriptions (catalogues of collections, metadata for specific objects) to more elaborate outputs (scans, 3D models, physical analyses, etc.).

In this context, we consider that cultural heritage institutions could engage into a dedicated course of actionalong the following lines:

- *Data delivery and services*: Each institution should be in a position to describe by which means and in which formats it can provide access to its digital resources. Compliance to international standards and good practices such as the DARIAH Seal of Approval could be a major asset in this respect;
- *Access, use and re-use*: The charter should offer the possibility for each institution to state its policy in this respect, and in particular to specify the access constraints (categories of users, fees) and the conditions (e.g. licences) under which a scholar can further publish any kind of content based upon material (documents, metadata) that it has initially made available;
- *Further hosting services*: The institution should be encouraged to notify when it can provide specific data hosting possibilities for amended or enriched content. By agreeing in the charter principles as a data-hosting infrastructure, it can contribute further to the improvement of data curation. In such a setting, even more digital surrogates can be curated in a coherent environment, with the notable advantage that this includes the connection to the corresponding primary data.

The benefits for the undersigning institutions are numerous. The institutions would potentially gain higher visibility for their assets and be able to boast their support to researchers, with the potential impact on the institutional support they would thus get. They would also have a direct feedback loop from the researchers themselves. The advantages of such a direct communication with the scholarly community include a better understanding of possible digitisation strategies in relation to higher societal interest of specific collections or of emerging research areas. Finally, they would be linked to potential digital hosting institutions (in the case that they are themselves unable to take on this role, for infrastructural reasons or otherwise) with whom they could then directly set up coherent collaboration schemes.

### Researchers

This charter aims at creating a direct relationship between scholars, CHIs and hosting infrastructures. This is why we contemplate that scholars could or even should sign in person, that is independently from the institution for which they are working at the time they sign the charter. However, we would welcome academic institutions (departments, universities, research institution or funding agencies) wishing to sign the charter and even make it a requirement for their members or the projects they fund or host.

As we cannot expect that scholars will themselves have the technical or human capacities to implement complex technical parameters and because we want to engage as large a group of people and institution as possible, we would limit the required commitment to the following aspects:
- *Compliance*: Scholars must in all cases certify that they will comply with the use and re-use conditions stated by the CHI signatories of the charter, in order to create a solid relationship of pledged trust between them. This concerns in particular citation and licencing conditions, that have to be respected without a single exception;
- *Contribution to the open dissemination process*: Material resulting from the scholar's work shall be further distributed in the most sophisticated form possible, for instance as source XML-TEI files, and in such conditions that further re-use is straightforward, in particular for other signatories of the charter;
- *Priority of deposit to the source*: When the cultural heritage institution from which the primary data is issued is offering this option, the scholars shall offer to deposit their own enrichments there first;

- *Priority to hosting institution which have signed the charter*: scholars should do their best to deposit their digital productions within undersigning institutions.

The charter could probably cover, maybe as an option, a larger scope of objects. Typically, the scope could also comprise authorities such as gazetteers, prosopographies or bibliographies compiled out of the primary sources used by the researchers. This has to be further discussed with all parties when finalising the charter.

The benefits of adhering to the data-re-use charter embrace different dimensions for the researcher. They include easier access to cultural heritage collections, fewer legal shilly-shallying when looking into the use and dissemination of digital material, and a far greater level of trustworthiness in the preservation of his/her production. The researcher hence has a maximal interest in being part of this endeavour. We should especially consider advocating the charter to researchers in such a way that they will contribute, through their own expressed preferences, to work with and alongside both CHIs and repositories, in order to highlight the direct benefits also to the latter of adherence to the charter.

## Data hosting infrastructures

Primary data can be hosted by CHIs or by Higher Education Institutions (HEIs) like universities, but they are in many cases curated by dedicated data hosting facilities as we have seen earlier in this paper. These institutions are equal partners in the Charter, alongside CHIs and scholars. They play a key role in guaranteeing the stability, the visibility and the long time availability of the primary data. The engagement we would expect for hosting facilities are clearly more technical and should ensure a concrete implementation of the CHI-researcher relation. We can think for instance of:
- Proper definition of the scope of the hosting facility in terms of types of data, accepted or required formats, additional descriptors (meta-data) attached to the data sets;
- Additional core services which are offered, in particular from the point of view of sustaining the content: long-term archiving, persistent identifiers, proper helpdesk, etc.;
- Access conditions to the archive: population being served (researchers, wider public), identification mechanisms, cost model (institution or project-based for instance).

Again, we can identify numerous advantages for the corresponding facility: better mapping of the service landscape, exchange of expertise, collection completion, critical mass of content (behind the scenes, easier identification by governmental agencies of hosting sectors that require long-term support/sustainability).

## Stakeholder inclusion

Devising and describing the content of such a charter will however be only a forerunner to ensuring that tall meaningful stakeholders are identified and are enabled to actively participate in the process. Beyond DARIAH members themselves, with the prospect of gaining national recognition for this charter, we do need to include various transnational initiatives in the library (LIBER, IFLA), archival (ICA) or scholarly domains (ADHO, EADH). Major aggregations such as Europeana would also need to be engaged at an early stage along with specific, flagship collecting institutions that could be potential early signatories. We should also make sure that smaller institutions such as regional historical archives for instance are included in the initial discussions so that they can also express their specific constraints, which may be quite different from more national insitutions.

Whatever the difficulty, we can anticipate of the huge workload that may be alleviated on all shoulders if this data re-use charter were set in place. While the implementation effort is expected to be minimal, the positive repercussions are expected to be nothing but marginal.

## Certification platforms

Putting what was described so far in perspective, let usturn again more specifically to the peoplethat we are actually here to serve as infrastructure providers, namely the researchers themselves. This turn of perspective is required for at least one major reason: because what we present as priority actions for DARIAH may be conversely considered by researchers as further constraints on their work. These constraints would affect the way they should describe their data, but also where they could deposit and host them, and finally also the conditions determining how other actors could re-use the content they have produced. There are many possible arguments that could demonstrate the added value of a more rigorous and channelled setting for data management, in particular in providing a wealth of data that would facilitate the life of humanities scholars at large and allow them to make many new discoveries, hypotheses and comparisons if the data is just there, and at hand.

Still, in the short term, we need to set up means to provide quick recognition for those who are spending a significant amount of scholarly time and effortinto designing and distributing high quality researchdata sets. This means that we need certification mechanisms which scholars could boast in relation to their data sets: an evaluation setting that should not be based solely on technical evaluations, but should also reflect possible appreciation by colleagues, departments, home institutions, research councils and funders, and not least, any formal national research assessment frameworks. There is no reason why this should not work in the same way and to the same extent as we have experienced in the context of similar regimes of research quality review, impact and assessment in the realm of thescholarly paper.

The underlying concept is not completely new. It has emerged and crystallized in the recent years around the notion of the data journal, instances of which have already appeared in various communities with, for instance, Geoscience Data Journal[25] or Earth System Science Data (ESSD)[26] in geosciences, the Biodiversity Data Journal[27] in biology or generic data journals such as CODATA's Data Science Journal[28].

The actual impetus towards providing appropriate environments here has attracted major publishing companies to launch similar initiatives. From high-profile publishers such as Nature Publishing Group with *Scientific Data* to opportunistic ones such as Hindawi publishing with its *Datasets* portfolio[29], not forgetting the profit-making "scholarly" association ACS with its *Journal of Chemical and Engineering Data*[30], the idea has already passed the stage of superficial interest, as it was still the case in the early2000s or even with forerunners such as the Journal of Astronomical Data (JAD)[31], which pioneered this approach in 1995.

---

[25] http://www.geosciencedata.com

[26] http://www.earth-system-science-data.net

[27] http://www.pensoft.net/journals/bdj/

[28] http://www.codata.org/dsj/index.html

[29] http://www.datasets.com/

Despite these examples from other disciplines, there still does not seem to be a similar momentum in the humanities. This means thatwe dispose – precisely now – of a unique opportunity to scope what humanities scholars would need and therefore, how DARIAH as an infrastructure would help establish or facilitate data journals in the humanitiesand contribute to their overall quality. This would not only encompass technical and logistical aspects, but also a critical assessment of the notion of "journal" itself; that is, what essentially constitutes the necessary form of publication for arts and humanities scholars who are increasingly sophisticated in their digital approaches to research. What is the content of a 'journal' in this sense and therefore, what apparatus is needed for, say, the tracking of citations relating to a much wider scope of material than bibliographic – programmes, software, algorithms, methodologies and data itself – how such ways of describing the impact also of this type of output integrate seamlessly with academic career structures?

First, the selection process in scholarly journals that has developed over the years is mainly based upon a post peer-review publication process that has until now prevented many potentially interesting results and studies to be published in the first place (see for instance [Jones, 2013]). Whereas this notion of selection originated in the, now digitally overcome,lack of available space in printed journals, this publication structureremains as a cultural relic in the scholarly ecoystem. If we want to ensure that all data sets in the humanities are actually made available, we should not aim at recreating print-world benchmarks and should think instead of crucially decoupling the actual publication (in the sense of "making public") from the assessment stage, by going towards post publication peer-review [Roberts, 1999].

Going even further, we should not just consider peer-review as an acceptation/rejection-based mechanism, but more as a ranking or certification service that may rate a data set along various editorial, technical and scholarly dimensions that may reflect how complex a data set can be as such, as well as how multifarious its post-publication environments of re-use might be. Indeed, if we consider for instance the publication of a corpus of transcribed videos corresponding to the recording of human interactions in context, where various phenomena (prosodic, referential, etc.) have been marked up, we want to offer the possibility to assess different aspects such as the conditions under which the experiment has been set up, the technical quality of the recordings, the appropriateness and precision of the transcription and annotations as well as the compliance of the corpus to existing international standards [Romary, 2015]. Such a certification process may also be carried out in other conditions than the traditional, not to say baroque, blind peer-review principles. Reviews may be made open to readers [Poeschl, 2012], as it is being implemented in the DH Commons Journal for instance. Reviewers may be selected by the submitter himself/herself, since s/he is probably the most likely person to know those who are appropriately most knowledgeable in his/her field. Finally, the data set can be left open for comment by the community, under clear authentication conditions to prevent spamming, and so forth.

---

[30] http://pubs.acs.org/journal/jceaax

[31] http://www.vub.ac.be/STER/JAD/JAD21/jad21.htm

The various editorial issues that we have outlined above demonstrate the actual need to experiment and to offer a flexible platform for future communities to develop certification and research assessment environments. Above all, we think that the wide availability of data sets and associated material should not rely on specific business models such as the one we have observed in APC-based "open access" journals. Such models, for instance, make the choice of the actual licence attached to the content depend upon the fee structures that the author is required to navigate and satisfy. Openness should not be hostage to fortune and we believe that we need, as a complement to having a network of publicly-owned data repositories in the humanities, (future) data journal platforms that should also remain in the realm of the public service[32].

This, however, is unlikely to be realised unless we can think of low-cost settings, which, in particular, do not duplicate existing data repository infrastructures. A possible answer is the establishment of overlay certification platforms which allow the data to be (first!) deposited in a given trusted data repository and whose reference will then be forwarded to the environment which adds a layer of peer- or community- review process, either of which may lead to a public assessment of the resource. This has been experimented in the context of the Episciences platform [Berthaud et alii, 2014]. This setting could serve as a freely available platform to set up experimental data journals. The steps towards establishing the necessary environments forsuch experiments can be itemized as follows:
- Identify a domain in the humanities where data production represents an important aspect of scholarly work;
- Identify criteria to assess the quality of such data production;
- Select a core of data repositories where such data are or could be hosted;
- Link these repositories to the EPisciences platform through its simple OAI/PMH interface;
- Motivate an enthusiastic editorial committee;
- Develop a satisfying workflow.

Not all aspects are equally simple to implement- we should not neglect human factors here.However, based on the arguments stated in the above, such initiatives are now necessary for the humanities. An infrastructure like DARIAH has to take on its role by helping to stabilize and promote as many of the necessary underlying components as possible.

## Ambitions for digital scholarship

As a whole, DARIAH is not just an administrative organisation coupled with a few European projects feeding its work plan. Since its very inception, it reflected upon the tremendous digital revolution that is occurring in the humanities and that has to be accompanied by a strong coordination of all actors, and it has developed its roadmap on this basis. Such a vision requires the continuous identification of the missing elements in the landscape, of the stumbling blocks that are still barriers for the humanities to be completely enabled to embrace the digital turn.

In the context of this article, we have highlighted a number of domains where we think that it is possible, thanks to the presence and capabilities of an infrastructure such as DARIAH on the side of humanities scholars, to initiate a deep change in the practices in digital scholarship. From very short-term aspects such as the repository landscape, to more sociologically complex proposals such as data journals, we think that it is necessary to propose and push a disruptive agenda, if only to make scholars themselves reflect upon their own digital future.

---

[32] This is indeed a major issue we see with such an initiative as : *Research Data Journal for the Humanities and Social Sciences*